\documentstyle[12pt]{article}
\newcommand{\be}{\begin{eqnarray}}
\newcommand{\ee}{\end{eqnarray}}

\begin{document}

\vspace{1cm}

\begin{center}

\LARGE{Radiative $\pi \rho$ and $\pi \omega$ transition form
factors in a light-front constituent quark model}\\

\vspace{1cm}

\large{ F. Cardarelli$^{(a)}$, I.L. Grach$^{(b)}$, I.M.
Narodetskii$^{(b)}$, \\ G. Salm\'e$^{(c)}$, S. Simula$^{(c)}$}\\

\vspace{0.5cm}

\normalsize{$^{(a)}$Istituto Nazionale di Fisica Nucleare, Sezione
Tor Vergata, Via della Ricerca Scientifica, I-00133 Roma, Italy\\
$^{(b)}$Institute of Theoretical and Experimental Physics, Moscow
117259, Russia\\ $^{(c)}$Istituto Nazionale di Fisica Nucleare,
Sezione Sanit\`{a}, Viale Regina Elena 299, I-00161 Roma, Italy}

\end{center}

\vspace{1cm}

\begin{abstract}

The form factors of the $\pi \rho$ and $\pi \omega$ radiative
transitions are evaluated within a light-front constituent quark
model, using for the first time the eigenfunctions of a light-front
mass operator reproducing the meson mass spectrum and including
phenomenological Dirac and Pauli quark form factors in the one-body
electromagnetic current operator. The sensitivity of the transition
form factors both to the meson wave functions and to the
constituent quark form factors is illustrated. It is shown that the
measurement of the $\pi \rho$ and $\pi \omega$ radiative
transitions could help in discriminating among various models of
the meson structure.

\end{abstract}

\vspace{1cm}

\newpage

\pagestyle{plain}

\indent The investigation of the radiative transition from the pion
to a vector meson ($\pi \gamma^* \rightarrow V$) could be of great
relevance for studying the meson structure and the mechanism of
quark confinement. Experimental information on $\pi \gamma^*
\rightarrow V$ transition could be provided by the extraction of
the pion-in-flight contribution from the electroproduction cross
section of vector mesons off the nucleon or light nuclei (cf.
experiments planned at high-intensity electron accelerator
facilities, like CEBAF \cite{PROPOSALS}). As far as the theoretical
side is concerned, it should be pointed out that for values of the
squared four-momentum transfer $Q^2 \sim$ few $(GeV/c)^2$ (i.e., in
the range of values of $Q^2$ accessible to present $CEBAF$
energies) hadron electromagnetic properties are expected to be
affected by the non-perturbative aspects of the QCD description of
exclusive processes (see, e.g., \cite{IL84}-\cite{BH94}).
Therefore, while waiting for a complete derivation of hadron form
factors from the fundamental theory, it is of interest to analyze
exclusive processes, like the radiative transition form factors of
mesons, adopting a relativistic constituent quark ($CQ$) model, in
which the meson is described as a bound state of a constituent $q
\bar{q}$ pair with all the other degrees of freedom being frozen in
the effective $CQ$ structure and effective $q \bar{q}$ interaction.
For an extensive discussion about i) the phenomenological success
of describing the pion as the hyperfine partner of the $\rho$ meson
and ii) the possibility to reconcile the CQ model with the
complexity of the QCD, see Refs. \cite{CI} and \cite{KGW},
respectively.

\indent The aim of this letter is to investigate the radiative
transitions $\pi^+ \gamma^* \rightarrow \rho^+$ and $\pi^0
\gamma^*  \rightarrow \omega$ for values of $Q^2$ up to few
$(GeV/c)^2$ within a light-front $CQ$ model. We make use of the
meson wave functions already adopted in \cite{CAR94}-\cite{CAR95}
for the investigation of the elastic $\pi$- and $\rho$-meson form
factors, i.e. we adopt the eigenfunctions of a light-front mass
operator, constructed from the effective $q \bar{q}$ Hamiltonian of
Ref. \cite{GI85} which reproduces the meson mass spectrum. In this
letter, the elastic pion and transition $\pi \rho$  ($\pi \omega$)
form factors are calculated, adopting an effective one-body
electromagnetic (e.m.) current operator which includes both Dirac
and Pauli form factors for the $CQ$'s (note that in Ref.
\cite{CAR94} the pion form factor has been investigated including
the $CQ$ Dirac form factor only). It is shown that the calculated
meson form factors are sensitive to the high-momentum tail,
generated in the wave function by the one-gluon-exchange ($OGE$)
part of the effective $q \bar{q}$ interaction, as well as to the
e.m. structure of the $CQ$'s. In particular, when the effects of
the configuration mixing in the meson wave functions are
considered, a non-vanishing value of the $CQ$ anomalous magnetic
moment is required in order to reproduce the experimental values of
the radiative decay widths of $\rho$ and $\omega$ mesons; moreover,
the comparison between our calculations and the experimental data
on pion form factor yields information on the possible size of
light $CQ$'s. Finally, our predictions are compared with those
obtained within various relativistic approaches, showing that the
measurement of the $\pi \rho$ and $\pi \omega$ radiative
transitions could help in discriminating among different models of
the meson structure.

\indent Let us remind that the matrix elements of the e.m. current
operator $\hat{I}_{\mu}$ for the transition $\pi \gamma^*
\rightarrow \rho$ ($\omega$) can be written as
 \be
    \langle P_{\pi}, 0 0 | \hat{I}_{\mu} | P_{\rho (\omega)}, 1
    \lambda_{\rho (\omega)} \rangle = F_{\pi \rho (\omega)}(Q^2)
    \epsilon_{\mu \nu \alpha \beta} e^{\nu}(\lambda_{\rho (\omega)})
    P_{\pi}^{\alpha} P_{\rho (\omega)}^{\beta}
    \label{1}
 \ee
where $Q^2 = - q \cdot q$ is the squared four-momentum transfer,
$F_{\pi \rho (\omega)}(Q^2)$ is the transition form factor,
$P_{\pi}$ and $P_{\rho (\omega)}$ are the four-momenta of the pion
and the $\rho$ ($\omega$) mesons, respectively, and $\lambda_{\rho
(\omega)}$ identifies the helicity state of the $\rho$ ($\omega$)
meson with polarization four-vector $e(\lambda_{\rho (\omega)})$.
In this letter, an effective one-body e.m. current operator
$\hat{I}_{\mu}$ is considered, viz.
 \be
    \hat{I}_{\mu} = \sum_{j=q, \bar{q}} \left [ F_1^{(j)}(Q^2)
    \gamma_{\mu} + F_2^{(j)}(Q^2) i \sigma_{\mu \nu} {q^{\nu}
    \over 2m_j} \right ]
    \label{2}
 \ee
where $F_1^{(q)}(Q^2)$ and $F_2^{(q)}(Q^2)$ are the Dirac and Pauli
quark form factors, normalized as $F_1^{(q)}(0) = e_q$ and
$F_2^{(q)}(0) = \kappa_q$, with $e_q$ and $\kappa_q$ being the $CQ$
charge and anomalous magnetic moment, respectively. It should be
pointed out that the two-body currents necessary for fulfilling
both the gauge and rotational invariances \cite{LIGHT-FRONT} have
not been taken into account in our calculations, where
phenomenological CQ form factors have been adopted.

\indent As in \cite{CAR94,CAR95}, the Poincar\'e-covariant state
vectors $| P_{\pi}, 0 0 \rangle$ and$| P_{\rho (\omega)}, 1
\lambda_{\rho (\omega)} \rangle$ are constructed using the
Hamiltonian light-front formalism \cite{LIGHT-FRONT}. Let us
briefly remind that the intrinsic light-front kinematical variables
are $\vec{k}_{\perp} = \vec{p}_{q \perp} - \xi \vec{P}_{\perp}$ and
$\xi = p^+_q / P^+$, where the subscript $\perp$ indicates the
projection perpendicular to the spin quantization axis, defined by
the vector $\hat{n}=(0,0,1)$, and the {\em plus} component of a
4-vector $p \equiv (p^0, \vec{p})$ is given by $p^+ = p^0 + \hat{n}
\cdot \vec{p}$; eventually, $\tilde{P} \equiv (P^+,
\vec{P}_{\perp}) = \tilde{p}_q + \tilde{p}_{\bar{q}}$ is the total
light-front momentum of the meson. In what follows, only the
$^3S_1$ channel of the $\rho$ ($\omega$) meson is considered, being
its $D$-wave component extremely small ($p_D \simeq 0.16 \%$) (see
\cite{CAR95}). Omitting for the sake of simplicity the colour
degrees of freedom, the requirement of Poincar\'e covariance for
the intrinsic wave function $\chi^j_{\mu} (\xi, \vec{k}_{\perp},
\nu \bar{\nu})$ of a meson with angular momentum $j$ and helicity
$\mu$ implies
 \be
   \chi^j_{\mu}(\xi, \vec{k}_{\perp}, \nu \bar{\nu}) = \sqrt{{M_0
   \over 16\pi \xi(1 - \xi)}} ~ R_{\mu}^j(\xi, \vec{k}_{\perp}, \nu
   \bar{\nu}) ~ w^{q \bar{q}}(k^2)
   \label{3}
 \ee
where $\nu, \bar{\nu}$ are the quark spin variables, $k^2 \equiv
k^2_{\perp} +  k^2_n$, $k_n \equiv (\xi - 1/2) M_0$, $M_0^2 =
(m_q^2 + k^2_{\perp}) / \xi$  $+ ~ (m_{\bar{q}}^2 + k^2_{\perp}) /
(1 - \xi)$ and the factor $R_{\mu}^j$ arises from the Melosh
rotations of the quark spins (see Ref. \cite{JAUS} for its explicit
expression). The wave function $\chi^j_{\mu} (\xi, \vec{k}_{\perp},
\nu \bar{\nu})$ is the eigenvector of a mass operator ${\cal{M}} =
M_0 + {\cal{V}}$, where the free mass operator $M_0$ acquires a
familiar form: $M_0 = \sqrt{m_q^2 + k^2} + \sqrt{m_{\bar{q}}^2 +
k^2}$, and $\cal{V}$ represents a Poincar\'e invariant interaction.
Therefore, the radial wave function $w^{q \bar{q}}(k^2)$ appearing
in Eq. (\ref{3}) is  an eigenfunction of the mass operator
$M=M_0~+~V$, obtained from the Melosh rotation of $\cal{M}$
\cite{CAR94,CAR95}. In particular $M_0$ commutes with the Melosh
rotation, while  $V$ is the Melosh-rotated interaction $\cal{V}$,
and it is i) independent upon the total momentum and the centre of
mass coordinates, ii)   rotationally invariant. Thus we have chosen
$w^{q \bar{q}}(k^2)$ as the eigenfunction of the effective $q
\bar{q}$ Hamiltonian, introduced by Godfrey and Isgur ($GI$)
\cite{GI85} for reproducing the meson mass spectra, viz.
 \be
    H_{q \bar{q}} ~ w^{q \bar{q}}(k^2) | j \mu \rangle & \equiv &
    \left [\sqrt{m_q^2 + k^2} + \sqrt{m_{\bar{q}}^2 + k^2} + V_{q
    \bar{q}} \right ] ~ w^{q \bar{q}}(k^2) | j \mu \rangle  =  M_{q
    \bar{q}} w^{q \bar{q}}(k^2) | j \mu \rangle
    \label{5}
 \ee
where $m_q$ ($m_{\bar{q}}$) is the constituent quark (antiquark)
mass, $M_{q \bar{q}}$ the mass of the meson, $| j \mu \rangle =
\sum_{\nu \bar{\nu}} ~ \langle {1 \over 2} \nu {1 \over 2}
\bar{\nu} | j \mu \rangle |{1 \over 2} \nu \rangle | {1 \over 2}
\bar{\nu} \rangle$ the equal-time quark-spin wave function and
$V_{q \bar{q}}$ the effective $q \bar{q}$  potential, composed by a
$OGE$ term (dominant at short separations) and a linear-confining
term (dominant at large separations). Three different forms of
$w^{q \bar{q}}(k^2)$ will be considered, namely $w^{q
\bar{q}}_{(conf)}$, $w^{q \bar{q}}_{(si)}$ and  $w^{q
\bar{q}}_{(GI)}$ corresponding to the solution of Eq. (\ref{5})
obtained using for $V_{q \bar{q}}$ only the linear confining term,
the spin-independent part and the full $GI$ interaction,
respectively. It should be pointed out that: i) in case of both
$w^{q \bar{q}}_{(conf)}$ and $w^{q \bar{q}}_{(si)}$ the meson mass
spectrum is badly reproduced, and ii) the $\pi$-meson ($^1S_0$
channel) and $\rho$-meson ($^3S_1$ channel) radial wave functions
differ only when the spin-spin component of the $q \bar{q}$
interaction is considered; this means that: $w^{\pi}_{(conf)} =
w^{\rho}_{(conf)} \equiv w_{(conf)}$, $w^{\pi}_{(si)} =
w^{\rho}_{(si)} \equiv w_{(si)}$ and $w^{\pi}_{(GI)} \neq
w^{\rho}_{(GI)}$. The four wave functions $w_{(conf)}$, $w_{(si)}$,
$w^{\pi}_{(GI)}$ and $w^{\rho}_{(GI)}$ have been already reported
in \cite{CAR95}, where it has been shown that both the central and
the spin-dependent components of the $OGE$ interaction strongly
affect the high-momentum tail of the $\pi$- and $\rho$-meson wave
functions. Herebelow, an ideal mixing in the vector sector is
assumed, because the effects of $\phi - \omega$ mixing are expected
to affect slightly the $\pi^0 \omega$ transition form factor (cf.
Ref. \cite{JAUS}). For the same reason, also the effects of the
$\rho^0-\omega$ mixing are neglected. Therefore, the same radial
wave function for the $\rho$ and $\omega$ mesons is adopted.
According to Ref. \cite{GI85}, the value $m \equiv m_q =
m_{\bar{q}} = 0.220 ~ GeV$ is adopted.

\indent Within the light-front formalism, the invariant hadron form
factors can be determined using only the matrix elements of the
component $I^+$ of the current operator; for space-like values of
the four-momentum transfer, we choose a frame where $q^+ = 0$, for
such a choice allows to suppress the contribution of the Z-graph
(pair creation from the vacuum) \cite{ZGRAPH}. Using Eqs.
(\ref{1}-\ref{3}) it can be checked that the matrix element
corresponding to $\lambda_{\rho (\omega)} = 0$ is vanishing; thus,
considering $\lambda_{\rho (\omega)} = 1$ and performing a
straightforward spin and flavour algebra \cite{CLO79}, one gets
 \be
    F_{\pi^+ \rho^+}(Q^2) ~ = ~ {1 \over 3} ~F_1^{(S)}(Q^2) ~
    H_1^{\pi \rho}(Q^2) ~ + ~ {1 \over 3} ~ F_2^{(S)}(Q^2) ~
    H_2^{\pi \rho}(Q^2)
    \label{6} \\
    F_{\pi^0 \omega}(Q^2) ~ = ~ F_1^{(V)}(Q^2) ~ H_1^{\pi
    \rho}(Q^2) ~ + ~ F_2^{(V)}(Q^2) ~ H_2^{\pi \rho}(Q^2) ~~~~
    \label{6bis}
 \ee
where $F_{1(2)}^{(S,V)}$ are the isoscalar and isovector parts of
the constituent $u$ and $d$ quark form factors, given by
$F_{\alpha}^{(S)}(Q^2)$ $\equiv 3 [F_{\alpha}^{(u)}(Q^2) +
F_{\alpha}^{(d)}(Q^2)]$ and $F_{\alpha}^{(V)}(Q^2)$ $\equiv
F_{\alpha}^{(u)}(Q^2) - F_{\alpha}^{(d)}(Q^2)$ (with $\alpha=1,2$
and $F_1^{(S)}(0) = F_1^{(V)}(0) = 1$) \footnote[1]{In the
derivation of Eqs. (\ref{6})-(\ref{6bis}) the relation
$F_{\alpha}^{(\bar{q})}(Q^2) = - F_{\alpha}^{(q)}(Q^2)$ has been
used.}. In Eqs. (\ref{6})-(\ref{6bis}) $H_1^{\pi \rho}(Q^2)$ and
$H_2^{\pi \rho}(Q^2)$ are body form factors, which depend on the
motion of the $CQ$'s inside the mesons and are given explicitely by
 \be
    H_1^{\pi \rho}(Q^2) = \int d\vec{k}_{\perp} d\xi ~ {\sqrt{M_0
    M'_0} \over 4 \xi (1 - \xi)} ~ {w^{\pi}({k'}^2) w^{\rho}(k^2)
    \over 4 \pi} ~ {m \lambda + 2 k_y^2 \over \lambda ~ \sqrt{m^2 +
    k_{\perp}^2} ~ \sqrt{m^2 + {k'}_{\perp}^2}} ~ 2 (1 - \xi)
   \label{8} \\
    H_2^{\pi \rho}(Q^2) = \int d\vec{k}_{\perp} d\xi ~ {\sqrt{M_0
    M'_0} \over 4 \xi (1 - \xi)} ~ {w^{\pi}({k'}^2) w^{\rho}(k^2)
    \over 4 \pi} ~ {\lambda (m^2 + \vec{k}_{\perp} \cdot
    \vec{k'}_{\perp}) - 2 M_0 (1 - \xi) k_y^2 \over m \lambda ~
    \sqrt{m^2 + k_{\perp}^2} ~ \sqrt{m^2 + {k'}_{\perp}^2}}
    \label{8bis}
 \ee
where $\lambda \equiv 2m + M_0$, $\vec{k'}_{\perp} \equiv
\vec{k}_{\perp} + (1 - \xi) \vec{q}_{\perp}$ and the $x$ axis is
chosen in the direction of $\vec{q}_{\perp}$ ($Q^2 \equiv
|\vec{q}_{\perp}|^2$). The expression (\ref{8}) for $H_1^{\pi
\rho}(Q^2)$ has been already derived in \cite{AO90}, whereas
$H_2^{\pi \rho}(Q^2)$ (Eq. (\ref{8bis})) is a new body form factor
related to the presence of the Pauli quark form factor in the
one-body e.m. current operator (\ref{2}). If the $u$ and $d$
constituent quarks have the same e.m. structure, which means
$F_{\alpha}^{(S)} = F_{\alpha}^{(V)}$, the $\pi \omega$ transition
form factor is three times the one corresponding to the $\pi \rho$
transition at any values of $Q^2$ and for any choice of the radial
wave function $w^{q \bar{q}}$. The body form factors $H_1^{\pi
\rho}$ and $H_2^{\pi \rho}$, calculated using the three radial wave
functions $w_{(conf)}^{q \bar{q}}$, $w_{(si)}^{q \bar{q}}$ and
$w_{(GI)}^{q \bar{q}}$, are shown in Fig. 1. It can be seen that
the effects of the spin-dependent part of the effective $q \bar{q}$
interaction are negligible, because the products $w_{(si)}^{(\pi)}
\cdot w_{(si)}^{(\rho)}$ and $w_{(GI)}^{(\pi)} \cdot
w_{(GI)}^{(\rho)}$ are very similar in a wide range of values of
the internal momentum (cf. Fig. 1 in \cite{CAR95}); in particular,
note that $H_2^{\pi \rho}$ is approximately two times larger than
$H_1^{\pi \rho}$.

 \indent The values of the transition form factors at $Q^2 = 0$
(the so-called transition magnetic moments) have been
experimentally determined from the radiative decay widths of the
$\rho$ and $\omega$ mesons, viz. $\mu_{\pi^+ \rho^+}^{exp} = 0.741
\pm 0.038 ~ (c/GeV)$ and $\mu_{\pi^0 \omega}^{exp} = 2.33 \pm 0.06
{}~ (c/GeV)$ \cite{PDG92}. From Eqs. (\ref{6})-(\ref{6bis}) one gets
$\mu_{\pi^+ \rho^+}$ $= [H_1^{\pi \rho}(0) + \kappa_S H_2^{\pi
\rho}(0)] / 3$ and $\mu_{\pi^0 \omega}$ $= H_1^{\pi \rho}(0) +
\kappa_V H_2^{\pi \rho}(0)$, where $\kappa_S \equiv 3(\kappa_u +
\kappa_d)$ and $\kappa_V \equiv \kappa_u - \kappa_d$ are the
isoscalar and isovector $u$ and $d$ quark anomalous magnetic
moments. Assuming $\kappa_S = \kappa_V = 0$ (which implies
$\mu_{\pi^0 \omega} = 3  \mu_{\pi^+ \rho^+}$), our results,
obtained using the radial wave functions $w_{(conf)}^{q \bar{q}}$,
$w_{(si)}^{q \bar{q}}$ and $w_{(GI)}^{q \bar{q}}$, are as follows:
$\mu_{\pi^+ \rho^+} = 0.833, 0.637, 0.561 ~ (c/GeV)$  respectively.
This means that: i) the value obtained using $w_{(conf)}^{q
\bar{q}}$ is close to the one reported in \cite{JAUS}, where a
simple Gaussian wave function was adopted; ii) the effects of the
configuration mixing, due to the $OGE$ interaction, lead to a $\sim
25\%$ underestimation of the experimental data. Then, the values of
$\kappa_S$ and $\kappa_V$ can be chosen in order to reproduce the
experimental values of the transition magnetic moments; in such a
way, using $w_{(GI)}^{q \bar{q}}$, one gets: $\kappa_S = 0.174 \pm
0.037$ and $\kappa_V = 0.208 \pm 0.019$, corresponding to $\kappa_u
= 0.133 \pm 0.016$ and $\kappa_d = - 0.075 \pm 0.016$. Note that,
within the quoted uncertainties, $\kappa_V \simeq \kappa_S$ due to
the fact that $\mu_{\pi^0 \omega}^{exp} \simeq 3 ~  \mu_{\pi^+
\rho^+}^{exp}$. Since non-vanishing quark anomalous magnetic
moments are required, the elastic form factor of the pion has to be
calculated including also the contributions arising from the Pauli
quark form factor not considered in \cite{CAR94}. Starting from the
relation $\langle {P'}_{\pi}, 0 0 | \hat{I}_{\mu} | P_{\pi}, 0 0
\rangle = F_{\pi}(Q^2) ~ (P + P')_{\mu}$ and using Eqs.
(\ref{2}-\ref{3}), one has
 \be
    F_{\pi}(Q^2) =  F_1^{(V)}(Q^2) ~ H_1^{\pi}(Q^2) ~ + ~
    F_2^{(V)}(Q^2) ~ H_2^{\pi}(Q^2)
    \label{10}
 \ee
with
 \be
    H_1^{\pi}(Q^2) = \int d\vec{k}_{\perp} d\xi ~ {\sqrt{M_0 M'_0}
    \over 4 \xi (1 - \xi)} ~ {w^{\pi}({k'}^2) w^{\pi}(k^2) \over 4
    \pi} {m^2 + \vec{k}_{\perp} \cdot \vec{k'}_{\perp} \over
    \sqrt{m^2 + k_{\perp}^2} ~ \sqrt{m^2 + {k'}_{\perp}^2}}
    \label{11} \\
    H_2^{\pi}(Q^2) = - {Q^2 \over 2} ~ \int d\vec{k}_{\perp} d\xi ~
    {\sqrt{M_0 M'_0} \over 4 \xi (1 - \xi)} ~ {w^{\pi}({k'}^2)
    w^{\pi}(k^2) \over 4 \pi} {1 - \xi \over \sqrt{m^2 +
    k_{\perp}^2} ~ \sqrt{m^2 + {k'}_{\perp}^2}}
    \label{11bis}
 \ee
The expression (\ref{11}) for $H_1^{\pi}(Q^2)$ has been obtained in
several papers (cf., e.g., \cite{CAR94}), whereas we stress that
$H_2^{\pi}(Q^2)$ (Eq. (\ref{11bis})) is a new body form factor
related to the presence of the Pauli quark form factor in the
one-body e.m. current operator (\ref{2}). The body form factors
$H_1^{\pi}$  and $H_2^{\pi}$, calculated using the three radial
wave functions $w_{(conf)}^{q \bar{q}}$, $w_{(si)}^{q \bar{q}}$ and
$w_{(GI)}^{q \bar{q}}$, are shown in Fig. 2. It can be seen that
the effects of the configuration mixing lead to a sharp increase of
the body form factors $H_1^{\pi}$  and $|H_2^{\pi}|$; in
particular, note that $|H_2^{\pi}|$ is substantially larger than
$H_1^{\pi}$ for $Q^2 > 1 ~ (GeV/c)^2$.  From Eqs.
(\ref{6})-(\ref{6bis}) and (\ref{10}) it turns out that both
$F_{\pi}$ and $F_{\pi^0 \omega}$ depends upon the isovector
combination $F_{\alpha}^{(V)}(Q^2)$ of $u$ and $d$ quark form
factors, whereas $F_{\pi^+ \rho^+}$ involves their isoscalar part
$F_{\alpha}^{(S)}(Q^2)$.

\indent We have evaluated the elastic pion form factor and the
transition $\pi \rho$ and $\pi \omega$ form factors, including in
the one-body e.m. current both Dirac and Pauli quark form factors
parametrized through simple monopole and dipole behaviours,
respectively, viz.
 \be
    F_1^{(q)}(Q^2) = {e_q \over 1 + (r_1^q)^2 Q^2 /6} ~~~~, ~~~~
    F_2^{(q)}(Q^2) = {\kappa_q \over (1 + (r_2^q)^2 Q^2 /12)^2}
    \label{12}
 \ee The parameters appearing in Eq. (\ref{12}) have been fixed as
follows. From Eq. (\ref{10}) the pion charge radius is given by
$<r^2>_{\pi} \equiv -6 {dF_{\pi} \over dQ^2}(0)$ $= <r_1^2>_{body}
+ \kappa_V <r_2^2>_{body}$ $ + (r_1^V)^2$, where
$<r_{\alpha}^2>_{body} \equiv -6 {dH_{\alpha}^{\pi} \over dQ^2}(0)$
and $(r_1^V)^2 \equiv e_u (r_1^u)^2 - e_d (r_1^d)^2$. Using the
value $\kappa_V = 0.208$, obtained from our previous analysis of
the transition magnetic moments $\mu_{\pi^+ \rho^+}$ and
$\mu_{\pi^0 \omega}$, and the experimental value $<r^2>_{\pi}^{exp}
= (0.660 \pm 0.024 ~ fm)^2$ \cite{PION-RADIUS}, one gets $r_1^V =
0.41 ~ fm$. If the e.m. structure of $u$ and $d$ $CQ$'s is assumed
to be the same, one has $r_1^u = r_1^d = r_1^V = 0.41 ~ fm$.
Finally, the value  $r_2^u = r_2^d = 0.52 ~ fm$ has been chosen in
order to get agreement in the whole range of existing pion data, as
it is shown in Fig. 3a, where our results are reported and compared
with the predictions of a simple Vector Meson Dominance ($VMD$)
model (i.e., $F_{\pi}^{VMD}(Q^2) = 1 / (1 + Q^2 / M_{\rho}^2)$,
with $M_{\rho}$ being the $\rho$-meson mass), the results of the
Bethe-Salpeter ($BS$) approach of Ref. \cite{IBG92} and those
obtained in Ref. \cite{NR} using $QCD$ sum rule techniques. It can
be seen that the differences among the theoretical calculations are
quite small, so that the existing pion data do not discriminate
among various models of the meson structure. Using the same $CQ$
form factors adopted for the description of the pion, we have
calculated the transition form factor $F_{\pi \rho}(Q^2)$. Our
results are reported in Fig. 3b, where they are compared with the
predictions of the simple $\rho$-pole $VMD$ model and the results
of Refs. \cite{IG93} ($BS$ approach) and \cite{BH94} ($QCD$ sum
rule technique). It can be seen that, unlike the case of the pion
form factor, the differences among various relativistic
calculations of the $\pi \rho$ transition form factor are quite
sizeable at $Q^2 > 1 ~ (GeV/c)^2$; therefore, the measurement of
$F_{\pi \rho}(Q^2)$ could help in discriminating among various
models of the meson structure. It is worth mentioning that the $\pi
\rho$ and $\pi \omega$ transition form factors play a relevant role
in the contribution of meson exchange currents ($MEC$) to the
deuteron elastic form factors (cf. Ref. \cite{IG93}); in
particular, the model dependence, clearly exhibited in Fig. 3(b) at
large $Q^2$, could produce sizable differences in the $MEC$
contribution to the deuteron form factors.

\indent It should be pointed out that the contributions of the
Pauli quark form factor to the elastic pion and transition $\pi
\rho$ ($\pi \omega$) form factors turn out to be at most $\sim 30
\%$. Then, from Eqs. (\ref{6})-(\ref{6bis}) it follows that the
ratio of the $\pi \omega$ to the $\pi \rho$ transition form factor
depends slightly on the meson wave function $w^{(q \bar{q})}$ and
is given mainly by the ratio of the isovector to the isoscalar
Dirac quark form factors, viz. $F_{\pi \omega}(Q^2) / F_{\pi
\rho}(Q^2)$ $\sim 3 ~ F_1^{(V)}(Q^2) / F_1^{(S)}(Q^2)$. Therefore,
the ratio of the $\pi \omega$ to the $\pi \rho$ transition form
factor is expected to be sensitive to possible differences between
$F_1^{(V)}(Q^2)$ and $F_1^{(S)}(Q^2)$, i.e. to possible differences
in the e.m. structure of the constituent $u$ and $d$ quarks. These
features are illustrated in Fig. 4, where our results, obtained
assuming various choices of the size parameters appearing in Eq.
(\ref{12}), are reported. It should be pointed out that each set of
values adopted for the $CQ$ size parameters satisfies the
constraints $e_u (r_1^u)^2 - e_d (r_1^d)^2 = (0.41 ~ fm)^2$ and
$e_u (r_2^u)^2 - e_d (r_2^d)^2 = (0.52 ~ fm)^2$, so that the
results of the corresponding calculations of the elastic pion form
factor remain in nice agreement with the existing pion data.

\indent In conclusion, the radiative transition $\pi^+ \gamma^* ->
\rho^+$ and $\pi^0 \gamma^* -> \omega$ have been analyzed within a
constituent quark model, in which the relativistic treatment of the
light constituents is achieved using the Hamiltonian light-front
formalism. In our approach the eigenfunctions of a light-front mass
operator, reproducing the meson mass spectrum, are adopted and both
Dirac and Pauli form factors are included in the one-body
electromagnetic current operator. When the effects of the
configuration mixing, generated in the meson wave function by the
one-gluon-exchange interaction, are considered, a non-vanishing
value of the constituent quark anomalous magnetic moment is
required in order to reproduce the experimental values of the
radiative decay widths of $\rho$ and $\omega$ mesons. The
contributions resulting both from Dirac and Pauli quark form factor
have been evaluated also in case of the pion and the comparison of
our results with existing pion data has provided information on the
size parameters of light constituent quarks. Moreover, it has been
shown that the ratio of the $\pi \omega$ to the $\pi \rho$
transition form factor is not affected by the choice of the meson
wave function, whereas it is sharply sensitive to possible
differences in the electromagnetic structure of $u$ and $d$
constituent quarks. Our predictions have been compared with the
results of various relativistic approaches, showing that the
measurements of $\pi \rho$ and $\pi \omega$ radiative transitions
could help in discriminating among different models of the meson
structure.

\vspace{0.5cm}

\indent Acknowledgements. We gratefully acknowledge E. Pace for
useful discussions and F. Gross for supplying us with the numerical
output of the calculations of Ref. \cite{IG93}.

\newpage

\vspace{0.5cm}

\begin{center}

{\bf Figure Captions}

\end{center}

\vspace{0.5cm}

Fig. 1. a) The body form factor $H_1^{\pi \rho}$ (Eq. (\ref{8}))
vs. $Q^2$. The dotted, dashed and solid lines correspond to the
calculations performed using the radial wave functions
$w_{(conf)}^{\pi}$, $w_{(si)}^{\pi}$ and $w_{(GI)}^{\pi}$, which
are the solutions of Eq. (\ref{5}) obtained using for $V_{q
\bar{q}}$ only the linear confining term, the spin-independent part
and the full $GI$ interaction \cite{GI85}, respectively. b) The
same as in a), but for $H_2^{\pi \rho}$ (Eq. (\ref{8bis})).

\vspace{0.5cm}

Fig. 2. a) The body form factor $H_1^{\pi}$ (Eq. (\ref{11})) vs.
$Q^2$. The dotted, dashed and solid lines correspond to the
calculations performed using the radial wave functions
$w_{(conf)}^{q \bar{q}}$, $w_{(si)}^{q \bar{q}}$ and $w_{(GI)}^{q
\bar{q}}$, respectively. b) The same as in a), but for $H_2^{\pi}$
(Eq. (\ref{11bis})).

\vspace{0.5cm}

Fig. 3. a) The elastic form factor of the pion (Eq. (\ref{10})),
times $Q^2$, vs. $Q^2$. Our results, obtained using
$w_{(GI)}^{\pi}$ and considering the $CQ$ form factors of Eq.
(\ref{12}) (with $r_1^u = r_1^d = 0.41 ~ fm$, $r_2^u = r_2^d = 0.52
{}~ fm$ and $\kappa_V = 0.208$), are represented by the solid line.
For comparison, the experimental data of Refs.
\cite{PION-RADIUS,PION-DATA} (cf. also \cite{CAR94}) are reported.
The dotted, dot-dashed and dashed lines correspond to the
predictions of a simple $VMD$ model ($\rho$-meson pole only), the
$BS$ approach of Ref. \cite{IBG92} and the $QCD$ sum rule technique
of Ref. \cite{NR}, respectively. b) The form factor of the
radiative transition $\pi^+ \gamma^* -> \rho^+$ (Eq. (\ref{6})),
times $Q^4$, vs. $Q^2$. The solid line correspond to our results,
obtained using $w_{(GI)}^{q \bar{q}}$ and considering the same
constituent quark form factors as in (a). The dotted, dot-dashed
and dashed lines correspond to the predictions of a simple $VMD$
model ($\rho$-meson pole only), the $BS$ approach of Ref.
\cite{IG93} and the $QCD$ sum rule technique of Ref. \cite{BH94},
respectively.

\vspace{0.5cm}

Fig. 4. The ratio of the $\pi^0 \omega$ to the $\pi^+ \rho^+$
radiative transition form factor (Eqs. (\ref{6})-(\ref{6bis})) vs.
$Q^2$. The dotted, dashed and solid lines correspond to the
calculations performed using $w_{(conf)}^{q \bar{q}}$, $w_{(si)}^{q
\bar{q}}$ and $w_{(GI)}^{q \bar{q}}$, respectively, and assuming
the same e.m. structure for the constituent $u$ and $d$ quarks
(viz., $r_1^u = r_1^d = 0.41 ~ fm$ and $r_2^u = r_2^d = 0.52 ~
fm$). The solid lines with open and full dots are the results of
the calculations obtained using $w_{(GI)}^{q \bar{q}}$ and assuming
different sizes for the constituent $u$ and $d$ quarks; namely, the
parameters appearing in Eq. (\ref{12}) are: $r_1^u = 0.37 ~ fm <
r_1^d = 0.49 ~ fm$, $r_2^u = 0.46 ~ fm < r_2^d = 0.61 ~ fm$ (open
dots), and $r_1^u = 0.44 ~ fm > r_1^d = 0.34 ~ fm$, $r_2^u = 0.56 ~
fm > r_2^d = 0.43 ~ fm$ (full dots). In all cases $\kappa_V =
0.208$ and $\kappa_S = 0.174$ (see text). Note that, for each set
of values of the $CQ$ size parameters, the corresponding
calculation of the elastic pion form factor is in nice agreement
with existing pion data.

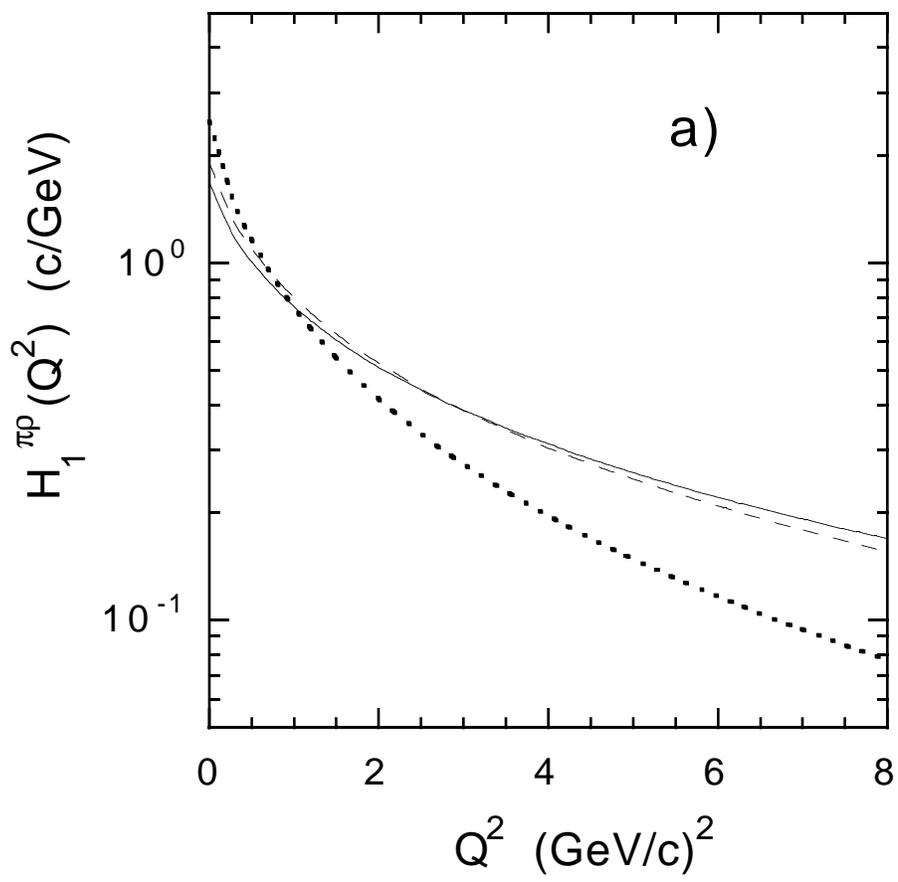
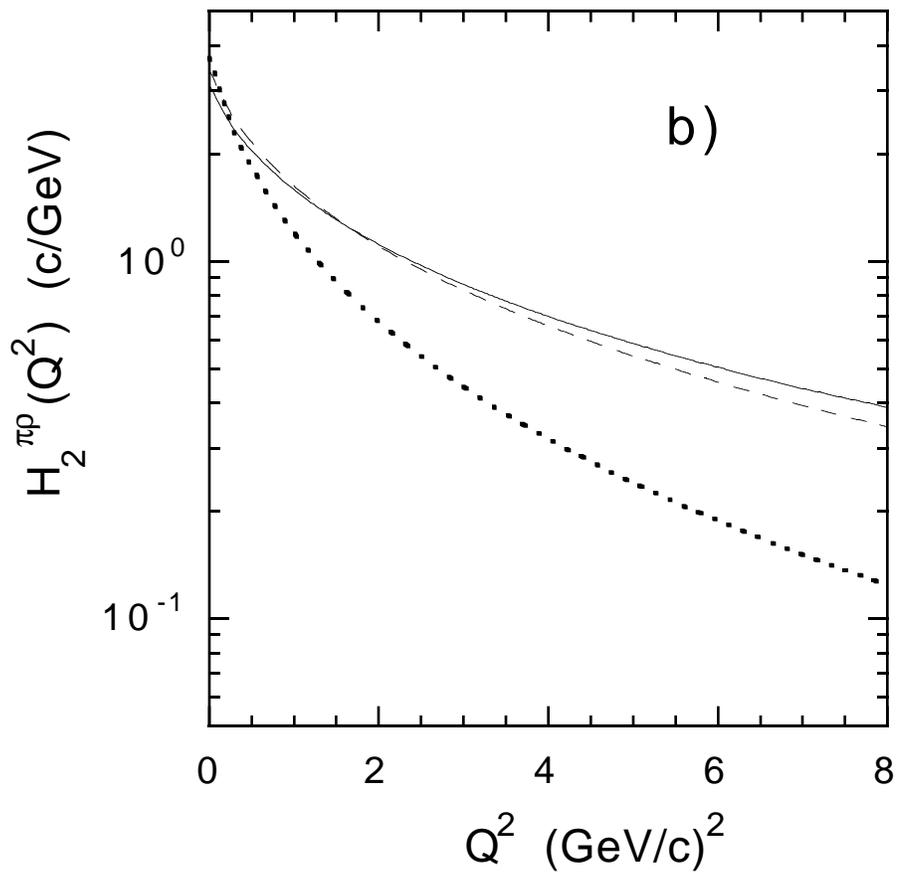
\end{document}